\documentclass[aps,pre,nofootinbib,amsmath,amsfonts,twocolumn]{revtex4}
\usepackage{hyperref}
\usepackage{graphicx,xcolor}
\usepackage{longtable}
\usepackage{subfigure}
\usepackage{xfrac}

\newcommand{\eqr}[1]{Eq.~\eqref{#1}}

\newcommand{\velocity}{{\rm{v}}}

\begin{document}

\title{Thermodynamics without ergodicity}
\author{V. L. Kulinskii}
\affiliation{Department of Theoretical Physics, Odessa National University, Dvoryanskaya 2, 65026 Odessa, Ukraine}
\author{K.~S.~Glavatskiy}
\affiliation{School of Civil Engineering, The University of Sydney, New South Wales 2006, Australia}

\begin{abstract}
We show that fundamental thermodynamic relations can be derived from deterministic mechanics for a non-ergodic system. This extend a similar derivation for ergodic systems and suggests that ergodicity should not be considered as a requirement for a system to exhibit a thermodynamic behavior. Our analysis emphasizes the role of adiabatic invariants in deterministic description and strengthens the  link between mechanics and thermodynamics. In particular, we argue that macroscopic thermodynamic behavior of a system is caused by the existence of different time scales in its deterministic microscopic evolution.
\end{abstract}

\maketitle



Since phenomenological thermodynamics received its microscopic justification by means of statistical mechanics, the question of whether there exists a similar justification by means deterministic mechanics has been open. Despite microscopic description (e.g. molecular simulations) of matter operates by dynamic variables (e.g. velocities) of individual particles, the macroscopic behavior of matter is determined by statistical distributions of the dynamic variables. The ability to use the statistical approach relies on the large number of particles which constitute a macroscopic body and on ergodicity of the system. In particular, this implies that the time average of a dynamic variable is equal to its phase space average. While the thermodynamic measurements represent essentially results of time averaging of some dynamic variable, ergodicity allows one to calculate those quantities rigorously using the methods of  statistical mechanics. Ergodicity has been proven for certain systems even with low degrees of freedom (e.g. Sinai's billiards \cite{Sinai1963}), while it still remains a hypothesis in general and there is no explicit criteria for ergodicity in terms of the Hamiltonian of the system.

The first attempt to build a deterministic description for a thermodynamic system dates back to the work of Hertz \cite{Hertz1910a, Hertz1910b} where he used purely mechanical notion of adiabatic invariance for a thermodynamic system. Later similar approach was developed by Krutkov \cite{Krutkov1922}, who had been inspired by Ehrenfest\cite{Ehrenfest1914}. They have proved adiabatic conservation of the phase space volume and showed that the entropy of the system is given by the logarithm of the phase space volume. This approach has not gained much attention though \cite{Adib2002}. Although based on mechanical arguments, their connection to thermodynamics used collective properties of the whole system (such as the phase space volume) rather than individual mechanical properties of the particles, therefore relying on the assumption of ergodicity.

In this paper we will show that ergodic hypothesis is neither necessary nor sufficient condition for a mechanical system to exhibit a thermodynamic behavior. In particular, we will show that for a  completely deterministic system it is possible to derive the fundamental relation, which combines the first and the second laws of thermodynamics for equilibrium processes
\begin{equation}\label{eq/00}
dE = T\,dS - p\,dV
\end{equation}
without any statistical assumptions. By doing this we also provide an entirely deterministic definition of the entropy, the quantity which has been believed to emerge only probabilistically. In particular,  we will show that the entropy of a deterministic system of $N$ particles in a $D$-dimensional volume is represented by the expression 
\begin{equation}\label{eq/101}
S_d = \log\frac{1}{N!}\Bigg(\sum_{i=1}^{D N}\,\Omega_i^2\Bigg)^{D N/2}
\end{equation}
where 
\begin{equation}\label{eq/102}
\Omega_i = \frac{1}{2\pi\hbar}\int{dp_i\,dq_i}
\end{equation}
is the phase space volume of a particle's single dimension with the coordinate $q_i$ and the momentum $p_i$. The expression under the logarithm can be interpreted as the volume of a $D N$-dimensional cube with the edge length equal to $\sqrt{\sum_{i=1}^{D N}\,\Omega_i^2}$. The deterministic entropy $S_d$ differs from the entropy of an ergodic system  $S_e$, which is equal to the logarithm of the phase space volume, by a constant independent of the system's temperature and the volume (but dependent on the number of particles), which does not affect the thermodynamic behavior of the system. Yet, $S_d$ provides a precise deterministic definition for the entropy without any reference to probabilistic nature of particles' velocities and positions distribution.

To introduce the notion of adiabatic invariance let us consider a single particle $i$ of mass $m$ which moves with the velocity $\velocity_i$ in a one-dimensional rigid box with the length $a$ and experiencing elastic collisions with the box walls. Let one of the walls move with a constant velocity $u << \velocity_i$, so the distance $a$ between the walls changes slowly and linearly with time. When the particle collides with the wall, the absolute value of its velocity changes by $2u$. The distance which is passed by the moving wall between two subsequent collisions of the particle and the non-moving wall is $\Delta a = u\,\left[a/(\velocity_i-u) + \{a+u\,a/(\velocity_i-u)\}/(\velocity_i-2u)\right]$. It it convenient to analyze the behavior of the abbreviated action $I_i$ of the particle \cite{ll1} 
\begin{equation}\label{eq/103}
I_i = \frac{1}{2\pi\hbar}\oint{m\,\velocity_i\,dq_i}
\end{equation}
where the integral is taken over the particle's trajectory between two subsequent collisions with the non-moving wall. Note, that the abbreviated action  defined by \eqr{eq/103} is equal to the phase space volume of the particle's single dimension defined by \eqr{eq/102}: $I_i = \Omega_i$.  Direct calculations show that $I_i = m\,\velocity_i\,a/(\pi\hbar)$ is independent of the box length and does not change when the wall moves: $I_i(a) = const$. Invariance of the abbreviated action under slow (so-called adiabatic) change of a parameter is called the adiabatic invariance. It has been studied extensively for dynamic systems \cite{LochakMeunier}. 

Consider now $N$ point particles of mass $m$ enclosed in a $D$-dimensional box. Without loss of generality we may assume that the box is cubic, so its edge length is equal to $a$ and the volume is $V = a^D$. The particles do not interact with each other, experiencing elastic collisions with the walls only. The velocity of each particle $\velocity_i$ remains the same between two consecutive collisions with either wall. This deterministic system is normally called the ideal gas and provides a basic representation of many thermodynamic systems. The total energy of the system is $E = \sum_{i=1}^{D N}\,{m_i\,\velocity_i^2/2}$. Expressing the velocity component of the particle $\velocity_i$ in terms of the abbreviated action component $I_i$, and introducing the ``kinetic action'' as
\begin{equation}\label{eq/104}
K^2 \equiv \sum_{i=1}^{D N}\,I_i^2
\end{equation}
we obtain that 
\begin{equation}\label{eq/105}
K(a) = \frac{a}{\pi\hbar}\,\sqrt{2m\,E(a)} = const
\end{equation}
is an invariant of the motion. When $a$ changes, so does the energy $E$ of the system, while $K$ remains constant. 

It follows from \eqr{eq/105} that for the process of slow variation of the system's volume (i.e. when $K=const$) the energy variation is $dE = -(2/D)\,E\,V\,dV$. Evaluating the derivative $p \equiv -(\partial E/  \partial V)_K$ we find that 
\begin{equation}\label{eq/106}
p\,V = \frac{2}{D}\,E
\end{equation}
which for $D=3$ is identical to the equation of state for the ideal gas. \eqr{eq/106} suggests therefore that it is natural to identify $p$ as the thermodynamic pressure. Furthermore, the process during which $K$ remains constant is the thermodynamic adiabatic process, during which the thermodynamic entropy of the gas remains constant. 

To make precise identification of the temperature and the entropy, we consider a general (nonadiabatic) process, during which $K$ may change. Differentiating both sides of \eqr{eq/105} we obtain
\begin{equation}\label{eq/107}
\frac{1}{2}\,\frac{dE}{E} + \frac{1}{D}\,\frac{dV}{V} = \frac{dK}{K}
\end{equation}
Note, that \eqr{eq/106} and \eqr{eq/107} are entirely deterministic equations, and we have not used any assumption about thermodynamic properties of the ideal gas. We next introduce the deterministic temperature $T_d$ and the deterministic entropy $S_d$ as
\begin{equation}\label{eq/108}
\begin{array}{rl}
\displaystyle \frac{1}{2}\,T_d\equiv& \displaystyle \frac{E}{D N}
\\\\
dS_d \equiv&\displaystyle D N\,\frac{dK}{K}
\end{array}
\end{equation}
Substituting these definitions in \eqr{eq/107} and using the above definition of pressure, we obtain \eqr{eq/00}. To be able to argue that we have derived a thermodynamic relation, we need to show that $T_d$ and $S_d$ have a thermodynamic meaning. This is indeed the case. In particular, for the ideal gas $E/(D N)$ is the average kinetic energy of the particle's degree of freedom. We know from classical statistical mechanics that this quantity is equal to the half of the temperature (measured in the energy units) of the system in the state of thermal equilibrium. Thus, $T_d$ can indeed be interpreted as the thermodynamic temperature $T$. Furthermore, substituting $K$ from \eqr{eq/105} in terms of $T$ and $V$ in expression for $S_d$ we obtain 
\begin{equation}\label{eq/109}
dS_d = N\,\left(\frac{D}{2}\,\frac{dT}{T} + \frac{dV}{V}\right)
\end{equation}
which for $D=3$ is identical to the expression for the entropy variation of the ideal gas with the temperature $T$ and volume $V$. It follows therefore that it is natural to identify $S_d$ as the thermodynamic entropy. 

Integrating the second of \eqr{eq/108} we obtain that $S_d = D N \log K + S_0$. Here $S_0$ is a constant of integration, which is independent of $T$ or $V$, but may depend on the number of particles. This is in agreement with the thermodynamic understanding of the entropy, which is defined up to a certain reference value. The integration constant may be obtained by demanding additivity of the deterministic entropy, which results in $S_0 = -\log{N!}$. The factor $N!$ reflects the fact that the total energy of the system is independent of how the particles are counted (index $i$ in \eqr{eq/104} is a running index), which is another statement of the  fact that the particles are indistinguishable. Combining all that  together, we obtain \eqr{eq/101} for the deterministic entropy $S_d$.

It is interesting to compare \eqr{eq/101} with the ergodic expression for the entropy $S_e = \log{(\Omega/N!)}$, where the phase space volume $\Omega$ accessible to the system is given by the product of the volume $D N$-dimensional sphere in the momentum space and the volume of the coordinate space \cite{Kubo}:
\begin{equation}\label{eq/110}
\Omega  = \frac{V^N}{(2\pi\hbar)^{D N}}\,(2m\,E)^{D N/2}\,\frac{\pi^{D N/2}}{(D N/2)!}
\end{equation}
where $\pi^{D N/2}/(D N/2)!$ is the volume of the $D N$-dimensional unit sphere. Substitution of \eqr{eq/105} in \eqr{eq/110} shows that the ergodic phase space volume
\begin{equation}\label{eq/111}
\Omega  = \Big(\frac{K}{2}\Big)^{D N}\,\frac{\pi^{D N/2}}{(D N/2)!}
\end{equation}
which means that $K$ defined by \eqr{eq/104} can be interpreted as the diameter of the $D N$-dimensional sphere of the ergodic phase volume of the system.

The above analysis shows that a deterministic mechanical system of $N$ point particles in a box reveals the thermodynamic behavior without any statistical assumptions. In particular, we  have shown that ergodicity is not required for a dynamic system to behave as a thermodynamic system. On the other hand, if a dynamic system is ergodic, as is assumed in any textbook on statistical mechanics, the approach presented in this paper is not valid and the thermodynamics is produced by the standard arguments of probabilistic statistical mechanics. 

This raises the question of the origin of the thermodynamic behavior. In particular, the entropy has always been believed to be a statistical quantity, which emerges as a collective property of the system consisting of many particles. Here we have shown that one can interpret the entropy as purely mechanical quantity, which is obtained from deterministic arguments. Furthermore, our derivation of the entropy does not use the so-called thermodynamic limit, which requires the number of particles to be very large. In fact, if the number of particles is small, we are still able to define the temperature and the entropy of the system according to \eqr{eq/108} and to derive the fundamental thermodynamic relation \eqref{eq/00}.

The mechanical analysis, which reveals the thermodynamic behavior is rigorous. The only implicit assumption which has been made is the ability of the system to undergo adiabatic transformations. That is the ability of the system's volume change slow enough, such that all the particles have managed to travel between the walls sufficient amount of times. In this case the trajectory of each particle is quasi-periodic and the integral over the closed trajectory in \eqr{eq/103} has a meaning. If the volume of the system changes fast enough, some particles' trajectories may not be closed, and the abbreviated action defined by \eqr{eq/103} would not be invariant. 

This suggests the following criterion for a mechanical system to reveal the thermodynamic behavior. The underlying dynamics of the system must be separable in a ``fast'' and a ``slow'' motion. In other words, there has to exist two distinct time scales: a ``microscopic'' time scale equal to a characteristic period of ``fast'' oscillations (within fixed external conditions) and a ``macroscopic'' time scale equal to a characteristic duration of the process (of changing the external conditions).

The separation of the time scales changes the nature of the dynamic variables, which is convenient to use to describe the evolution of the system. A standard mechanical description uses the instantaneous position and the velocity of a particles. This tradition has been adopted by statistical mechanics, which uses the instantaneous positions and momenta of the particles $q(t), p(t)$. However, this set of dynamic variables is not the only possible one: a set of canonical variables called  the action and the angle, $I(t), \phi(t)$  may be used equivalently in mechanics. The action variable is defined by \eqr{eq/103} as the integral over the particle's trajectory and the angle variable is defined as the phase along that trajectory. In the case of quasi-periodic motion the equations of motion in terms of the  action-angle variables are exceptionally simple: $dI/dt = 0$, $d\phi/dt = dH/dI$. Since the action variable is constant, the dynamic evolution of the particle is entirely accounted by the angle variable $\phi(t)$. The separation of the time scales results essentially in separation of the variables: fast microscopic oscillations are described  by the angle variable, while slow macroscopic process is described by the action variable.

The separation of the variables, in turn, changes the focus of the dynamic evolution from instantaneous coordinates to entire trajectories. In the case of a non-ergodic system every particle has its own (quasi-periodic) trajectory.  The motion of the particle along that trajectory is governed by the angle variable and represents fast microscopic oscillations and is irrelevant for the macroscopic behavior. In contrast, the evolution of the trajectory as a whole is governed by the action variable and represents slow macroscopic process.

As it is known, the thermodynamic description of a system allows one to greatly reduce the number of the relevant variables. Instead of using the $2 N D$ variables representing the particles' positions and momenta, one can describe a closed system with only two variables, the temperature and the volume. This reduction of variables is associated with the reduction of information necessary to describe the system. This allows one to interpret the entropy of the system as a measure of the information reduction. The deterministic definition of the entropy allows us to identify the exact reason for this reduction.

As it is evident from \eqr{eq/101}, the deterministic entropy is independent of the angle variables. This means that motion of a particle along its trajectory is irrelevant for the macroscopic behavior. What matters -- is the entire trajectory as a whole, which is described by a single number represented by the action variable $I$. Even though the instantaneous evolution of the coordinates $q(t), p(t)$ matters for the mechanical description of the system, it is irrelevant for the thermodynamic description. This is the first reason for the emergence of the entropy. The second reason is that from the thermodynamic perspective the particles are indistinguishable. While for the microscopic mechanical description it might be important to trace the evolution of every particle, for the thermodynamic behavior the relevant quantities are the total energy and the total volume.

V.K. is grateful to Mr.~Konstantin Yun for financial support of the research.
%

\end{document}